# Fundamental picture of the conduction mechanism in solid-state polymer electrolytes revealed by terahertz spectroscopy

Johanna Weidelt[1], Jijeesh Ravi Nair[2], Diddo Diddens[2], Wentao Zhang[1], Felix Pfeiffer[2], Tiago de Oliveira Schneider[3], Markus Meinert[3], Tomoki Hiraoka[1], Linda Nesterov[1], Masoud Baghernejad[2*], Dmitry Turchinovich[1], and Hassan A. Hafez[1**]

[1]Fakultät für Physik, Universität Bielefeld, Universitätsstr. 25, 33615 Bielefeld, Germany

[2]Helmholtz-Institute Münster, IEK-12, Forschungszentrum Jülich GmbH, Corrensstrasse 46, 48149 Münster, Germany

[3]Department of Electrical Engineering and Information Technology, Technical University of Darmstadt, Merckstraße 25, 64283 Darmstadt, Germany

**ABSTRACT:** Solid polymer electrolytes (SPEs) based on cross-linked poly(ethylene oxide) (PEO) encompassing lithium salts have gained significant attention as separators in solid-state lithium metal batteries. Here, we employ terahertz time-domain spectroscopy (THz-TDS), as a noninvasive contact-free technique, to investigate the conduction properties of these cross-linked SPEs and unravel their dependencies on the added lithium salt and the sample temperature. The obtained THz conductivity spectra are dominated by THz absorption bands, which we attribute to resonant vibrations within the polymer matrix of the electrolyte. By careful application of Lorentz model, the conductivity spectra have been analyzed, and the relevant polymer vibration modes have been quantitatively assessed. Calculations based on the density functional theory (DFT) were performed to elucidate the possible microscopic mechanisms of these resonant vibrations. This study sheds light on the relevance of polymer matrix vibrations validating the hopping transport of lithium ions in SPEs which ultimately leads to the technologically relevant ionic conduction in the solid-state polymer-based electrolytes.

## INTRODUCTION

Solid-state electrolytes hold a great promise for enhancing the energy density of lithium batteries without compromising safety [1–4]. Among such electrolytes, solid polymer electrolytes (SPEs) based on poly(ethylene oxide) (PEO) are highly attractive due to their flexibility, easy fabrication, mechanical strength, low toxicity, low cost, and good thermal stability[1,5–7]. Added lithium salts dissociate in the polymer matrix and the resulting lithium ions (Li+) coordinate with the partially negatively charged oxygen atoms of the PEO ether groups, as illustrated schematically in Figure 1. Enabled by the flexibly vibrating polymer matrix, the Li+ transport occurs via hopping along and between the polymer chains[1,5,8–13]. Additionally, the Li+ can also be transported alongside a mobile polymer segment while remaining attached[2,14,15]. Yet, the resulting ionic conductivity (typically ~$10^{-4}$ S cm$^{-1}$) of unmodified SPEs is relatively low, as compared with that of liquid electrolytes [1,2]. A multi-salt approach has thus been proposed to improve the conduction properties of SPEs [2,16,17]. With a mixture of two lithium salts, an enhancement of the resulting ionic conductivity by at least 62 % compared to the respective single-salt system has been realized[2]. However, further efforts to improve the conduction properties of SPEs require deeper insights into the associated molecular and ionic dynamics in these electrolytes and a better understanding of the salt-mixing effects on the polymer chain dynamics that enable the Li+ transport.

Terahertz time-domain spectroscopy (THz-TDS) is a powerful contact-free, all-optical non-destructive method to probe low-frequency IR-active vibrations, charge transport and dielectric relaxations in many material systems, including inorganic solids and polymeric materials[18–26]. This is accomplished by analyzing the complex-valued transmission spectra obtained from measuring time-dependent electric fields of THz light pulses transmitted through the material of interest[19,27]. Here, we apply THz-TDS, within the frequency range of 0.1-7 THz (3-233.5 cm$^{-1}$), to study the conduction properties of PEO-based electrolytes with samples containing Li+ salts, including lithium difluoro(oxalato)borate (LiDFOB), lithium bis(trifluoromethanesulfonyl)imide (LiTFSI), and their mixture, at controlled temperatures, ranging from 10 to 300 K. Compared with the bare polymer without any salts, highly distinguishable THz conductivity spectra were obtained from the electrolyte-samples with salts, which are found to depend strongly on the content and type of the added salt for the same oxide-to-Li$^+$-ratio. The experimental THz conductivity spectra are well reproduced by a Lorentzian model accounting for temperature and salt dependent resonant vibration modes. The assignment of these modes was established by performing additional calculations based on the density functional theory (DFT). The DFT calculations demonstrate that the twisting motion of the PEO backbone dominates in the bare polymer, while additional modes emerge and dominate in the salt-containing electrolytes due to the vibration of the entire crown-ether-like coordination shell around the Li$^+$ ion. Such low-frequency vibrations detected by THz spectroscopy are plausibly correlated with the electrically measured ionic conductivity, indicating that the observed vibrational modes serve the Li$^+$ ion transport.

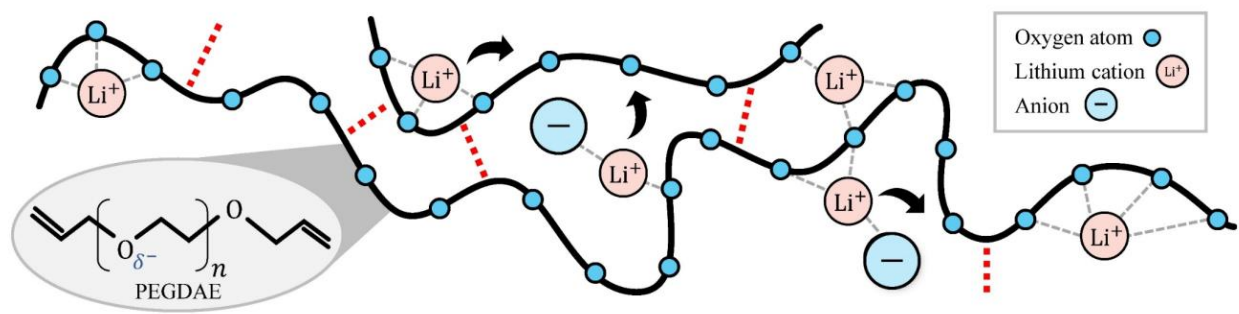

**Fig. 1.** Schematic illustration of the hopping charge transport in PEO-based electrolytes: intra- and interchain hopping, either direct or mediated by the salt anions. Black lines represent the PEGDAE chains with the small blue dots corresponding to the comprised oxygen atoms (the chemical structure of PEGDAE is shown on the bottom left). The chemical structure of the anions depends on the comprised salt. Red dashed lines indicate cross-linking.

## SAMPLES AND EXPERIMENTS

**Cross-linked polymer electrolytes.** The samples of electrolytes used in this study are based on telechelic allyl-terminated PEO chains [poly(ethylene glycol) diallyl ether, PEGDAE]. These polymer chains are cross-linked via UV-induced free radical photopolymerization. Details of this fabrication process are reported elsewhere[2]. Two groups of samples, all prepared under the same conditions, were used in two experiments of THz-TDS, as we shall discuss shortly. Each group contained four samples described as follows: a first sample of bare cross-linked PEGDAE (x-PEGDAE), named X(0), a second sample of xPEGDAE containing 25 wt% LiTFSI, named XS1, a third sample of xPEGDAE containing 14 wt% LiDFOB, named XS2, and finally xPEGDAE containing a mixture of 20 wt% LiTFSI and 3 wt% LiDFOB, named X(S1+S2). All the chosen salt concentrations correspond to an ethylene oxide-to-Li$^+$-ratio of 17, yielding the maximum dc conductivity within the dual salt electrolyte (~$2.15 \times 10^{-4}$ S cm$^{-1}$ at 60 °C, as reported in [2]). Further information about the samples is provided in Note 1 and Figure S1 in the Supporting Information.

**Experiment 1.** THz-TDS with a table-top broadband THz spectrometer was performed with the samples of one group at room temperature (300 K). The broadband spectrometer is based on a spintronic emitter consisting of a magnetic Ta/CoFeB/Pt heterostructure, pumped by Ti:sapphire laser pulses at a repetition rate of 1 kHz. The broadband THz emission is thus enabled by laser-induced inverse spin Hall effect in the heterostructure[28,29]. The generated THz pulses exhibit a spectral amplitude that spans reliably from 0.1 THz to 8 THz, which were detected via electro-optic sampling, using a 200 μm-thick GaP crystal.

**Experiment 2.** THz-TDS at cryogenic temperatures down to 10 K were performed with the samples of the other group. In this experiment, we used a commercial THz spectrometer (TeraFlash Pro from Toptica Photonics AG) operating at a much higher repetition rate of 100 MHz. The free-standing samples were mounted in a closed-cycle helium cryostat that allowed for the measurements at low temperatures. The cryostat is equipped with two quartz optical windows transparent (~ 80% field transmission) for THz frequencies up to only 2.5 THz, while higher frequencies were not accessible in this experiment due to high reflection and absorption losses caused by the quartz windows. Given also that the electrolyte samples exhibit a high THz absorption at higher frequencies, it was thus sensible to use a spectrometer with a much higher repetition rate in this experiment, in order to maintain a high signal-to-noise ratio of the acquired data. The samples were first cooled down to 10 K and the transmitted THz fields have been measured at different fixed temperatures on the ascending mode of temperature change.

In each experiment, control measurements were performed through a sample-free aperture, serving as a reference. To avoid the THz absorption by the water molecules of humid air, all the measurements in the two experiments were conducted in dry nitrogen environment. The analysis of the experimental THz-TDS data to extract the THz conductivities of the samples follows a common procedure [19,22,24,30], which is described in some detail in Note 2 of the Supporting Information.

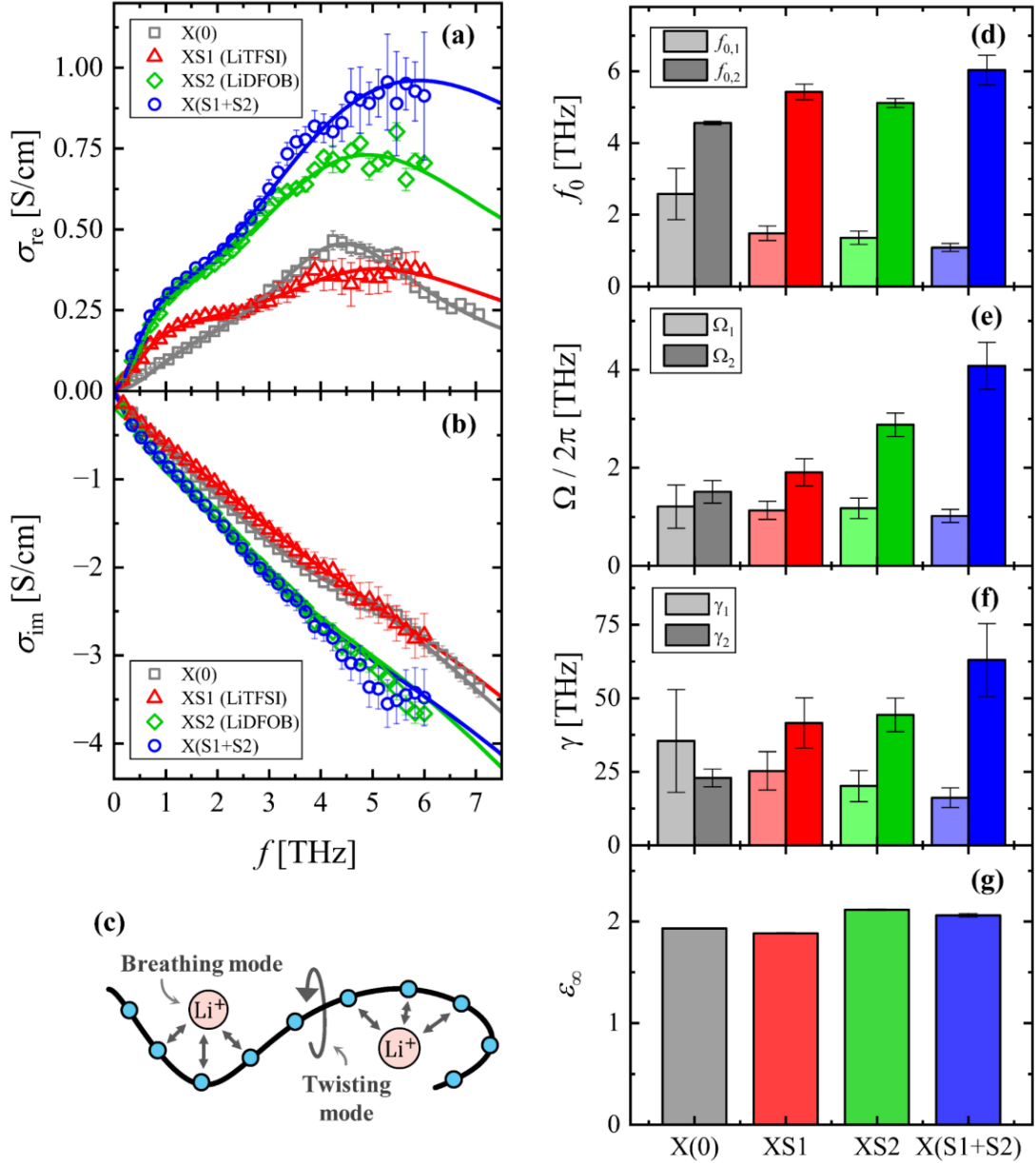

**Fig. 2.** (a) Real and (b) imaginary part of the frequency-dependent conductivity spectra of bare cross-linked polymer (X), single salt electrolytes (XS1 (LiTFSI) and XS2 (LiDFOB)) and dual salt electrolyte (X(S1+S2) (LiTFSI+LiDFOB)). Symbols for data points obtained by the broadband spintronic emitter, and solid lines to the Lorentzian fit. The error-bars are derived from the thickness uncertainty of the samples along with the standard error of the mean of the data acquired in repeated measurements. (c) Schematic of the dynamics associated with the resonance frequencies: the twisting motion of the polymer backbone and the breathing movement of the crown-like oxygen shell around the $Li^+$. (d) The two resonance frequencies obtained by fitting the data for each sample, respectively. (e) The sample-dependent fitting damping rates of the first and second resonant modes and (f) the associated oscillator strengths. (g) The high-frequency dielectric constant obtained by the fit.

**RESULTS AND DISCUSSION**

**Broadband THz conductivity at room temperature.** Figures 2a,b show the real and imaginary parts of the THz conductivities, respectively, for the samples of the first group, represented by symbols for the data and solid lines for the fitting based on the Lorentz oscillator model. For all the samples, the real-valued conductivity increases with frequency and reaches a peak at a certain frequency that varies slightly from sample to another. The lack of data points at frequencies above 6 THz for the conductivity of the electrolyte samples in Figures 2a,b is due to their large THz absorption, so that the THz transmission reaches the experiment's noise floor at these frequencies (see Notes 2 and 3 of the Supporting Information for the data analysis and specifying the trustful data points based on the dynamic range of our experiments).

As apparent in Figure 2a, the addition of salt results in a spectral modification of the THz conductivity, as compared with that of the bare polymer X(0). The salt-induced change of the conductivity spectra is distinctive upon the type of the added salt. Comparing the two single salt electrolytes, we find that XS2 clearly exhibits a larger real-valued THz conductivity over the entire frequency range shown in Figure 2a as well as a larger ionic conductivity (see Note 7 of the Supporting information) compared to XS1. We note here that the ionic conductivity values account for the mobility of the comprised ionic species within the electrolyte, while the frequency-dependent conductivity values caught by the THz spectroscopy account for the movement of bound charges, i.e., the vibration modes of the polymer matrix which might serve the electronically measured ionic mobility. Indeed, for the salt concentrations regarded here, XS2 (LiDFOB) provides a higher ionic conductivity compared to XS1 (LiTFSI) because of the higher charge mobilization of Li$^+$ in the presence of DFOB anion (weak interaction of Li$^+$ with PEO chains contributed by DFOB anion). This may also be due to a larger amplitude of the molecular vibrations of the polymer chains and in turn enhanced Li$^+$ hopping mobility.

Remarkably, the addition of a small amount of LiDFOB (with 3-5 wt%) to a LiTFSI-based electrolyte (with 20 wt%) results in enhancement of the real-valued THz conductivity as well as the ionic conductivity (see Note 7 in the Supporting Information) which are found to approach those of XS2 with 14 wt% LiDFOB. In Figure 2a, we see that the THz conductivity of X(S1+S2) exceeds that of XS2, especially at the higher frequencies above 3 THz. The mixing of salts likely increases the efficiency of the Li$^+$ transport via anion mediation, supported by suppressed ionic aggregate formation and increased amorphicity of the PEO matrix compared to a system purely based on LiTFSI[2,16,31]. In electrical measurements, this enhanced ionic mobility is observed to be the strongest for dual salt electrolytes containing a mixture of 20 wt% LiTFSI and 5 wt% LiDFOB (see Figure S7b in the Supporting Information). However, also for the dual salt electrolyte X(S1+S2) considered in the present study, which contains 3 wt% LiDFOB and 20 wt% LiTFSI, the THz-TDS reveals an enhanced real-valued THz conductivity, as compared with the LiTFSI-based single-salt electrolyte, as shown in Figure 2a. Despite this undoubtable enhancement of the conductivity obtained by the mixed salt approach, the deviation between the extracted THz conductivities of the two samples XS2 and X(S1+S2) is also subject to small (~4%) local thickness variations in the samples, which will be discussed further when discussing the results of Experiment 2 below. Yet, the clearer enhanced real THz conductivity of X(S1+S2) in the high frequency range above 3 THz stems possibly from enhanced polymer chain mobility due to the increased amorphicity of X(S1+S2) compared to XS2, as indicated previously[2].

To obtain further insights into the molecular dynamics and the Li$^+$ transport mechanisms in these electrolytes at higher frequencies, we examined different physical models to reproduce the obtained conductivity spectra (see Note 4 of the Supporting Information for further information). The best matching with the experimental data is obtained by applying a pure Lorentzian model featuring resonant THz absorption modes characteristic of polymer chain vibrations, as shown in Figures 2a,b. The exploited complex-valued Lorentzian conductivity function is given by[19,27,32,33]

$$\tilde{\sigma} = -i\varepsilon_0\omega(\varepsilon_\infty - 1) - \sum_j \frac{i\varepsilon_0\omega\Omega_j^2}{(\omega_{0,j}^2 - \omega^2) - i\gamma_j\omega} \quad (1)$$

where $\varepsilon_0$ is the vacuum permittivity, $\varepsilon_\infty$ is the high-frequency dielectric constant, and, for each mode of order $j$, $\Omega$ is the vibration amplitude (or oscillator strength), $\omega_0$ is the resonance angular frequency, $\gamma$ is the damping rate, and the sum is applied over the possible modes within the frequency range of our experiments. The model fitting is represented by the solid lines in Figures 2a,b, considering two resonant vibrational modes (i.e. the summation in Eq. (1) is taken over two terms with j=1,2) and the corresponding fitting parameters are depicted in Figures 2d-g. We note here that extrapolation of the experimental data, and accordingly the fitting conductivity function, do not show any significant contribution of a dc conductivity. This is not surprising because the revealed THz conductivities shown in Figure 2a are at least five orders of magnitude larger than the typical ionic conductivity of ~$10^{-5}$ S cm$^{-1}$ for these electrolytes. Hence, the addition of a Drude-like contribution including such a low dc conductivity has a negligible effect on the fitting quality provided by Eq. (1), and was therefore ignored. It is also worth noting here that Debye relaxation model is commonly used to describe conductivity and dielectric spectra of polymers and electrolytes with data obtained from experiments with narrower THz bandwidths extending to a maximum frequency of 3 THz[23,24,34]. However, our choice of the Lorentzian fitting function of Eq. (1) is motivated by the spectral features of the experimental conductivity spectra obtained by using the broadband THz spectrometer, and is further supported by density functional theory (DFT) calculations that will be discussed in the next section. Our approach to justify the exploited fitting model is also consistent with earlier studies on polymers and electrolytes[23,26,34–36].

Concerning the conductivity spectra of the bare polymer sample X(0) shown in Figures 2a,b, the best fit using Eq. (1) is achieved by including two broad resonant modes, one centered at $f_{0,1} = \omega_0/2\pi = $ 2.58 THz with a small intensity and a more pronounced mode centered at $f_{0,2} = \omega_0/2\pi =$4.56 THz. We attribute these modes to the twisting motion of the polymer backbone, as illustrated schematically in Figure 2c. The evident broadening of the modes, rather than sharp resonant peaks, can be attributed to the amorphicity of the polymer. Our fitting algorithm, employed to simulate the experimental data using Eq. (1), has directly provided these two central frequencies as well as the magnitudes of the other fitting parameters, as demonstrated in Figures 2d-g. We note that similar resonant absorption modes have been previously observed in transmission spectra of pure non-cross-linked crystalline PEO without allyl-termination[34,36]. Regarding the samples containing salts, these resonant modes persist with variations in the associated fitting parameters, i.e. their central frequencies, intensities and spectral widths, as shown in Figures 2d-g. The low-frequency mode at $f_{0,1}$, which lies within the most reliable part of the data, seemingly exhibits a red-shift and spectral narrowing with the addition of salts. These effects vary ascendingly with the sample sequence XS1, XS2, and X(S1+S2). In contrast, the second mode at $f_{0,2}$ seems to exhibit an opposite behavior, i.e. a blue-shift and spectral broadening, and additionally increasing oscillator strength (or peak intensity) with the aforementioned salt sequence. However, the lack of reliable experimental data at higher frequencies, especially for the X(S1+S2) sample, makes the evaluation of the salt effects on this mode uncertain. Therefore, we sought deeper insights into relevant molecular and ionic dynamics via performing DFT calculations.

**DFT calculations.** To rationalize the inherent vibration modes of the PEO backbone and how they are affected by the presence of salt, DFT calculations have been carried out using the B3LYP functional[37] and the 6-31+G(d,p) basis set[38,39]. Various clusters composed of PEO, Li$^+$ and/or anions (TFSI/DFOB) have been employed in the calculations. The intermolecular environment outside these clusters has been treated by the implicit SMD solvation [40]. All calculations have been performed with the Gaussian 16 package[41]. The geometries have been optimized and subsequently their vibrational frequencies have been computed. Although a possible phonon activity (e.g. in partially crystalline PEO domains) would not be captured by this cluster approach, it will nonetheless contain information about the characteristic backbone vibrations.

The clusters shown in Figure 3 revealed local vibrational modes with significant intensity at around 3.3-3.8 THz, mainly corresponding to a twisting motion of the PEO backbone. When salt is added, these vibrations persist, but their intensity is reduced as the coordination to Li$^+$ somewhat limits the twisting motion of the PEO backbone. However, additional vibrations at about 4.7-5.1 THz emerge, corresponding to the vibration of the entire crown-ether-like coordination shell around Li$^+$ (typically composed of 5-6 PEO oxygen atoms). It should be kept in mind that the clusters in Figure 3 reflect simplified model systems, while the vibrations observed in a real polymer melt reflect an average over a multitude of local PEO conformations and Li$^+$ coordination environments. Nevertheless, on a qualitative level, the additional vibrations of the crown-ether-like coordination Li$^+$ shell observed in the DFT calculations rationalize the emergence of a blue-shifted, enhanced and broadened second adsorption peak in the electrolyte samples compared to the bare polymer. This is also reflected by Figure 3, in which the results of the DFT calculations for the possible resonant polymer vibrations are plotted. Sampling the entire spectra would involve averaging over numerous clusters or the use of MD simulations[42], which is beyond the scope of our present study. Similarly, we refrain from performing DFT calculations for mixed-salt clusters because it would require an average over numerous possible arrangements of the two different anions.

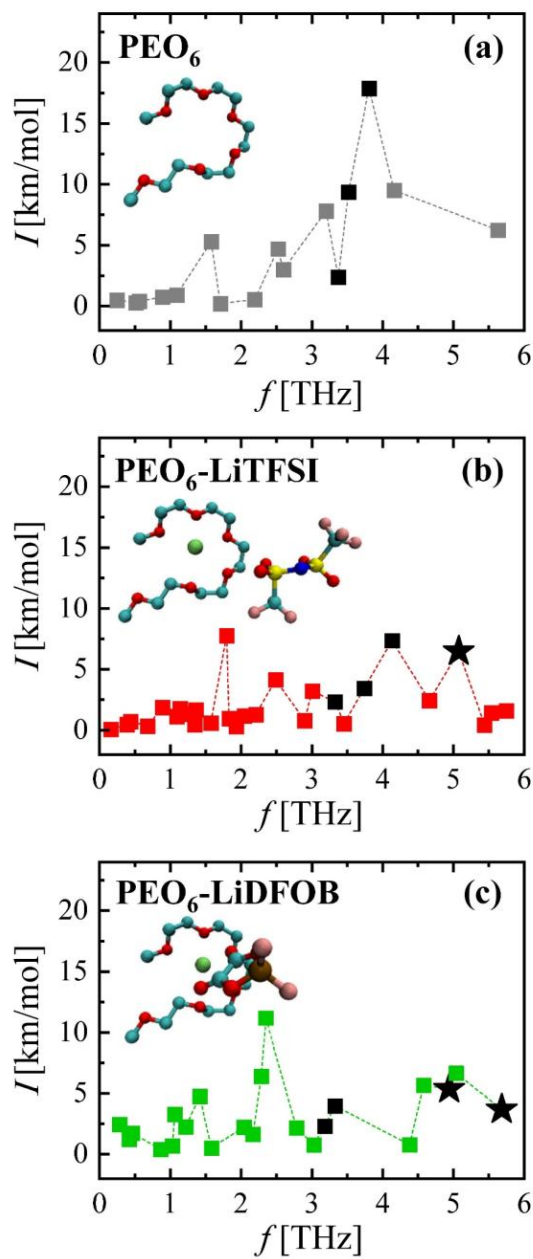

**Fig. 3:** DFT predicted resonances with snapshots depicting the associated clusters being a PEO hexamer (a) and Li$^+$ together with the anions TFSI (b) or DFOB (c). Hydrogen atoms have been omitted for clarity. The frequencies associated with the twisting motion of the polymer backbone are marked by black squares while the ones linked to the breathing movement of the crown-ether-like oxygen shell around the Li$^+$ are represented by black stars.

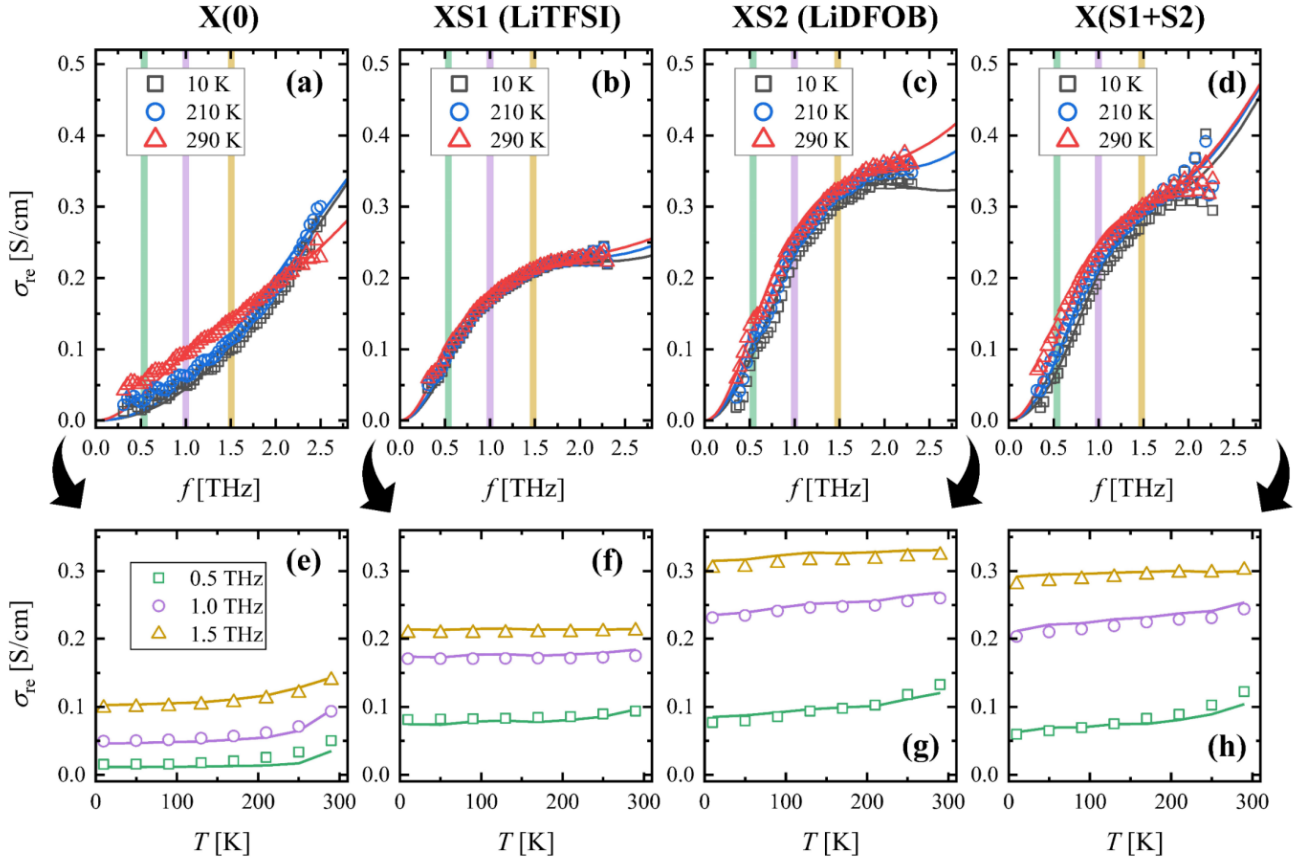

**Fig. 4: Temperature-dependent real-valued conductivity.** For each sample, the real-valued conductivity spectra as function of frequency at different temperatures are shown in the respective upper panel, while the temperature dependence of the conductivity at selective frequencies indicated by the vertical colored lines is shown in the respective lower panel. [(a), (e)] for X(0), [(b), (f)] for XS1, [(c), (g)] for XS2, and [(d), (h)] for X(S1+S2). Symbols for the data and solid lines for the Lorentz model fitting.

**Temperature dependence.** Figure 4 shows the spectra of the real-valued conductivity obtained from Experiment 2 at different cryogenic temperatures ranging from 290 K to 10 K. The upper row of graphs shows the frequency dependent THz conductivity spectra of the four samples at different temperatures, while the lower row shows the temperature dependence of conductivity at three selected frequencies, 0.5 THz, 1 THz and 1.5 THz. We observe a small increase of the conductivity with the increase of temperature, with a slight change of the slope when the temperature increases above 230 K, which is more visible in the cases of the bare polymer and the dual salt electrolyte. This behavior agrees qualitatively with the trend observed through temperature-dependent electrical measurements using dc and low-frequency electric fields[2]. At higher frequencies exceeding 2 THz, the temperature dependency of the samples X(0) and X(S1+S2) is reversed, i.e. the THz conductivity decreases slightly with the increase of temperature, while the single-salt samples do not show any remarkable change. The slight change of the slope around 230 K, as seen in the lower row of Figure 4, might plausibly be attributed to phase transition around the glass transition temperature $T_g$[2]. At this temperature, the amorphous phase of the polymer matrix switches from glassy to rubbery state as the temperature increases, allowing for segmental motion of polymer chains at room temperature[1,43,44]. Thus, the THz-TDS exhibits potential to serve as a convenient tool to probe this transition, as also suggested in earlier studies[45].

Consistent with the results from the broad band measurement presented in Figure 2, Figure 4 shows that the real-THz conductivity of XS1 is the lowest compared to the other electrolytes XS2 and X(S1+S2), thereby resembling the situation with the ionic conductivities measured electrically. Also in agreement with Experiment 1, the spectra obtained from XS2 and X(S2+S2) converge to each other in the regarded spectral window from 0.3 to 2.3 THz. However, slight deviations between the THz conductivities of these two electrolytes can be realized when comparing Figure 2a to Figures 4c,d,g,h, which we attribute to local thickness variations of ~4% at maximum realized in each sample (see also Note 1 in the Supporting Information regarding the sample thicknesses).

**Table 1:** Comparison of ionic conductivity, $\sigma_{ion}$, measured by electrical impedance spectroscopy (EIS) and spectrally weighted real-valued THz conductivity $\sigma_{sw}$, at the room temperature. Here $\sigma_{sw,nb}$ is calculated over the narrowband $0.3 \leq f \leq 2.3$ THz and is averaged for the data of the two experiments shown in Figures 2a and 4b-d , while $\sigma_{sw,bb}$ is calculated over the broadband $0.3 \leq f \leq 6$ THz from the data shown in Figures 2a.

| Sample | XS1 | XS2 | X(S1+S2) |
|---|---|---|---|
| Ionic conductivity, $\sigma_{ion}$/S cm$^{-1}$, at 293 K | 3.80×10$^{-6}$ | 9.76×10$^{-6}$ | 7.5×10$^{-6}$ |
| Average spectrally weighted THz conductivity, $\sigma_{sw,nb}$ /S cm$^{-1}$, over the frequency range $0.3 \leq f \leq 2.3$ THz | (18.58±0.76) ×10$^{-2}$ | (28.9±0.8) ×10$^{-2}$ | (28.87±0.93) ×10$^{-2}$ |
| Ratio $\sigma_{ion}/\sigma_{sw,nb}$ | (1.99±0.03) ×10$^{-5}$ | (3.38±0.10)×10$^{-5}$ | (2.60±0.08)×10$^{-5}$ |
| Spectrally weighted THz conductivity, $\sigma_{sw,bb}$/S cm$^{-1}$, over the frequency range $0.3 \leq f \leq 6$ THz | (28.30±2.88) ×10$^{-2}$ | (53.36±1.33) ×10$^{-2}$ | (62.64±4.33) ×10$^{-2}$ |
| Ratio $\sigma_{ion}/\sigma_{sw,bb}$ | (1.34±0.14)×10$^{-5}$ | (1.83±0.05)×10$^{-5}$ | (1.20±0.08)×10$^{-5}$ |

As an attempt to correlate the ionic conductivity to the obtained THz conductivity of these electrolytes quantitatively, we developed an approach based on calculating the ratio between the ionic conductivity, $\sigma_{ion}$, and the spectrally weighted THz real conductivity, $\sigma_{sw} = \frac{1}{f_b - f_a}\int_{f_a}^{f_b} \sigma_{re}(f)\,\mathrm{d}f$, where $f_a$ and $f_b$ represent THz frequency limits. As indicated in Table 1, we see distinctive ratios for the three types of investigated electrolytes, which also depend on the THz spectral range considered in the calculation. This makes the one-to-one correlation challenging, especially when considering the intense resonant band around 5 THz captured by the broadband spectroscopy of Experiment 1. However, the ratios lie within the range 1.2-1.8×10$^{-7}$, and represent qualitatively enhancement of the ionic conductivity as the THz vibrational activity of the electrolyte increases. Yet, a more reliable quantitative correlation requires further investigation, encompassing additional electrolytes with varying salt concentrations, which may be considered in future works.

We realize from the upper row graphs of Figure 4 that the temperature dependent behavior of the THz conductivities of the samples originates plausibly from spectral broadening upon temperature increase. We adapted the free-fitting parameters of the Lorentzian conductivity of Eq. (1) to obtain the best fit with the experimental data (the solid lines in Figure 4). We note here that the weak temperature dependence of the conductivities as well as the lack of data at frequencies above 2.5 THz make the assessment of the temperature dependence of the fitting parameters of Eq. (1) challenging. However, the best fitting was obtained by adapting the low-frequency resonance $f_{0,1}$, the damping rates $\gamma_1$ and $\gamma_2$ and the high-frequency dielectric constant $\varepsilon_\infty$ as slightly varying functions of temperature, as shown in Figure 5, while the other fitting parameters of Eq. (1) remained temperature-independent. The values chosen for those temperature-independent parameters were obtained by fitting the narrow-band data from Experiment 2 at room-temperature together with the broadband data from Experiment 1 for each sample (the associated parameter values are listed in Table S5 in the Supporting Information). We realize a red-shift of the low-frequency resonant mode at $f_{0,1}$, indicating a mode softening, with the increase of temperature, as shown in Figure 5a. The associated damping rate $\gamma_1$ does not show significant dependence on temperature, except for a sudden increase with temperature above 250 K for the case of the bare polymer sample X(0), as shown in Figure 5b. In contrast, the temperature dependence of the damping rate $\gamma_2$ associated with the dominant resonance mode at $f_{0,2}$ is highly pronounced. The increase of $\gamma_2$ (i.e. the mode broadening) with the increase of temperature can be attributed to increasing disorder (amorphicity) of the polymer matrix upon heating[39]. As such, THz spectroscopy could serve as a contact-free and non-destructive tool to probe molecular vibrational dynamics that enable the charge transport within SPEs, consequently providing information about the technologically relevant ionic conductivity.

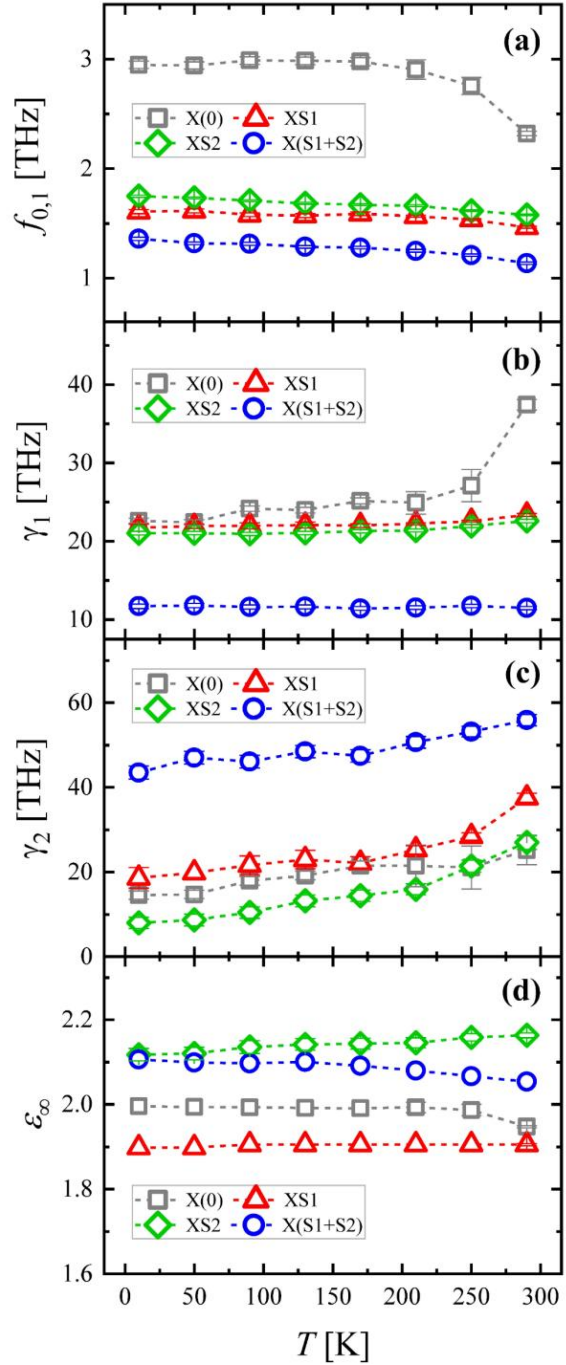

Fig. 5: Temperature- and sample dependent Lorentzian fit parameters: (a) the low-frequency resonance $f_{0,1}$, (b) the associated damping rate $\gamma_1$, (c) the damping rate $\gamma_2$ belonging to the high frequency resonance $f_{0,2}$, and (d) high-frequency dielectric constant. The lines are guide-to-the-eye.

## CONCLUSION

In this work, we studied the conduction properties of solid-state PEO-based electrolytes at THz frequencies. The extracted THz conductivity spectra are well reproduced by a pure Lorentzian model featuring resonant THz vibration modes whose characteristics vary from one sample to another depending on the added Li salt. In general, salt addition causes an increased vibrational activity, especially when LiDFOB is added or the two salts are mixed, which is likely serving for the increase of the Li+ mobility and the electrolyte dc conductivity reported in [2]. The revealed THz resonant modes were found to exhibit a small spectral band-broadening, and a red-shift and increase of the low-frequency side of the peaking THz conductivity band upon heating from 10 to 290 K. We attribute this behavior to an increase of the disorder of the polymer segments (increase of amorphicity) and mode softening upon heating. Additionally, the observed increase of THz conductivity shows two distinct trends around a temperature close to the glass temperature of the PEO polymer. With these findings, we hope that the present study motivates continuous efforts to optimize the conduction properties of this class of electrolytes, and to utilize THz spectroscopy in their characterization.

## ASSOCIATED CONTENT

**Supporting Information**. Two groups of samples; Time-domain THz signals; data analysis; complex-valued frequency dependent refractive index; models of conductivity; aging and environmental effects; and dc conductivity measurements. This material is available free of charge via the Internet at http://pubs.acs.org.

## AUTHOR INFORMATION

**Authors and Affiliations**

Johanna Weidelt, Wentao Zhang, Tomoki Hiraoka, Linda Nesterov, Dmitry Turchinovich, and Hassan A. Hafez - Fakultät für Physik, Universität Bielefeld, Universitätsstr. 25, 33615 Bielefeld, Germany

Jijeesh Ravi Nair, Diddo Diddens, Felix Pfeiffer, and Masoud Baghernejad – *Helmholtz Institute Münster, Corrensstraße 46, 48149 Münster, Germany*

Tiago de Oliveira Schneider, Markus Meinert - *Department of Electrical Engineering and Information Technology, Technical University of Darmstadt, Merckstraße 25, 64283 Darmstadt, Germany*

**Corresponding Authors**

*Masoud Baghernejad: b.masoud@fz-juelich.de
**Hassan A. Hafez: hafez@physik.uni-bielefeld.de

**Author Contributions**

H. Hafez, M. Baghernejad and D. Turchinovich conceived the idea and supervised the project. M. Baghernejad, J. Nair and F. Pfeiffer prepared the samples, performed the dc conductivity measurements and the supplementary Raman spectroscopy. J. Weidelt, H. Hafez, W. Zhang and T. Hiraoka performed the THz-TDS measurements. T. de Oliveira Schneider and M. Meinert provided the THz spintronic emitter. J. Weidelt, L. Nesterov and H. Hafez performed the data analysis and the fitting. D. Diddens performed the DFT calculations. J. Weidelt and H. Hafez wrote the manuscript with contributions from M. Baghernejad, J. Nair and D. Diddens. All co-authors discussed the results and commented on the manuscript.

**Funding Sources**

The authors acknowledge the financial support from the European Union's Horizon 2020 research and innovation programme (Grant Agreement No. 964735 EXTREME-IR), Deutsche Forschungsgemeinschaft (DFG) within Project No. 468501411-SPP2314 INTEGRATECH under the framework of the priority programme SPP2314-INTEREST and within Project No. 518575758 HIGHSPINTERA, and German Federal Ministry for Education and Research within the project EFoBatt (grant No. 13XP5129).

## ACKNOWLEDGMENT

The THz emitter was fabricated with facilities of Prof. G. Reiss at Bielefeld University in the Center for Spinelectronic Materials and Devices. The authors would like to acknowledge fruitful discussions with Nicolas Beermann, Alexander Stroh and Pascal Strathkötter.

## ABBREVIATIONS

PEO, poly(ethylene oxide); THz, terahertz; THz-TDS, THz time-domain spectroscopy; SPE, solid polymer electrolyte; SPE, solid polymer electrolyte; $Li^+$, lithium (ion), DFT, density functional theory.

Supporting Information for

# Fundamental Picture of the Conduction Mechanism in Solid-State Polymer Electrolytes Revealed by Terahertz Spectroscopy

Johanna Weidelt[1], Jijeesh Ravi Nair[2], Diddo Diddens[2], Wentao Zhang[1], Felix Pfeiffer[2], Tiago de Oliveira Schneider[3], Markus Meinert[3], Tomoki Hiraoka[1], Linda Nesterov[1], Masoud Baghernejad[2*], Dmitry Turchinovich[1], and Hassan A. Hafez[1**]

[1]Fakultät für Physik, Universität Bielefeld, Universitätsstr. 25, 33615 Bielefeld, Germany

[2]Helmholtz-Institute Münster, IEK-12, Forschungszentrum Jülich GmbH, Corrensstrasse 46, 48149 Münster, Germany

[3]Department of Electrical Engineering and Information Technology, Technical University of Darmstadt, Merckstraße 25, 64283 Darmstadt, Germany

**Note 1. Samples**

The two groups of samples used in this work are described as shown in Fig. S1. Group 1 was used in the first experiment with broadband THz signals, while Group 2 was used in the second experiment investigating the temperature dependence of the electrolytes' properties using a narrowband THz spectrometer. The thicknesses of the samples were measured mechanically by using a Vernier caliper (the results taken from 10 measurements for each sample).

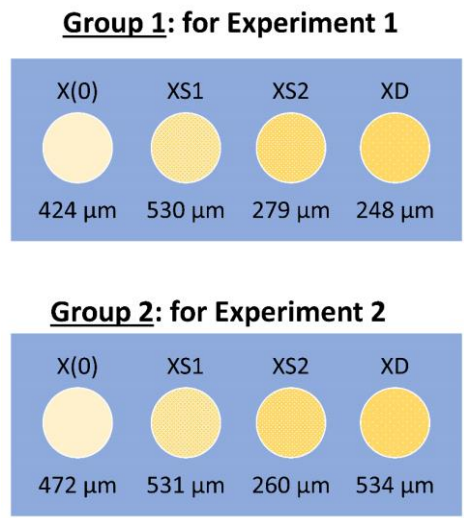

**Figure S1:** Two groups of samples used in the two experiments. The given thicknesses are of an uncertainty of $\pm 10$ μm.

**Note 2. Data from Experiment 1**

Fig. S2 (a) shows the measured THz signals transmitted through each sample at room temperature, as well as the reference signal measured in the absence of samples. For each pair of sample and reference signals, we define the time-dependent THz electric fields as $E_{sam}(t)$ and $E_{ref}(t)$, respectively. With the Fourier-transformation of these fields, $\tilde{E}_{sam}(\omega)$ and $\tilde{E}_{ref}(\omega)$, we obtain the amplitude spectra shown in Fig. S2(b), and the corresponding complex-valued frequency-dependent transmission, $\tilde{T}(\omega) = \tilde{E}_{sam}(\omega)/\tilde{E}_{ref}(\omega)$, which is related to the complex refractive index $\tilde{n} = n + i\kappa$ of the sample material through the approximate relation[1,2]

$$\tilde{T}(\omega) = \frac{\tilde{E}_{\text{sam}}(\omega)}{\tilde{E}_{\text{ref}}(\omega)} \approx \frac{4n}{(n+1)^2} \exp\left[-\frac{\kappa \omega d}{c}\right] \cdot \exp\left[i(n-1)\frac{\omega d}{c}\right] \quad (1)$$

where $\omega$ is the angular frequency, $d$ is the sample thickness, and $c$ is the speed of light in vacuum. This approximate relation is valid with our samples because the Fabry-Perot etalon multi-reflections within the samples were found negligible. The validity of this approximation has been further cross-checked versus other approaches such as the "Nelly"-algorithm[3] and an iterative ansatz accounting for the Fabry-Perot etalon effects[1,4]. The complex-valued dielectric function and conductivity are given by $\tilde{\varepsilon}(\omega) = \varepsilon_r(\omega) + i\varepsilon_i(\omega) = \tilde{n}^2(\omega)$ and $\tilde{\sigma}(\omega) = -i\varepsilon_0 \omega \cdot [\tilde{\varepsilon}(\omega) - 1]$, respectively[1,5–8], where $\varepsilon_0$ is the vacuum permittivity. The absorption coefficient is given by $\alpha(\omega) = 2\omega\kappa/c$. Figs. S2 (c-f) show the obtained spectra of the real refractive index, $n$, the absorption coefficient, $\alpha$, and the real- and imaginary parts of the dielectric function, $\varepsilon_r$ and $\varepsilon_i$, respectively, for the samples of Group 1. The trustful data points were determined for each sample by calculating the maximum measurable absorption coefficient[9]

$$\alpha_{max}(\omega) = \frac{2}{d} \ln\left[DR \frac{4n}{(n+1)^2}\right] \quad (2)$$

where $DR$ is the dynamic range of the THz-TDS experiment. For the trustful data points, the absorption coefficient must not exceed the corresponding $\alpha_{max}$. This resulted in the cut-off frequencies, the upper limit of the trustful spectral range, indicated in Fig. S2(d).

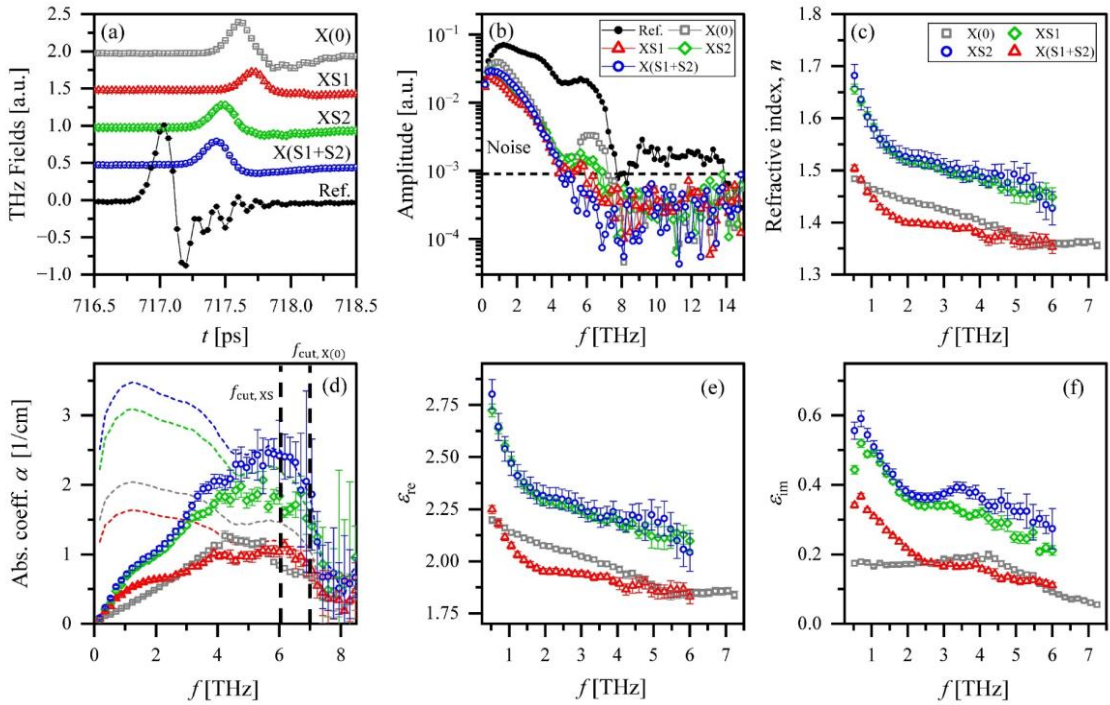

**Figure S2:** Data from Experiment 1. (a) the temporal waveforms of the THz signals transmitted through the samples and the reference signal transmitted through a sample-free aperture, (b) the corresponding amplitude spectra obtained by the fast Fourier transformation of the temporal signals of (a), (c) the real-valued refractive index. The arrows indicate the upper limit of the trustful spectral window for the trustful data for each sample. (d) the absorption coefficient, symbols for samples and dashed-lines for $\alpha_{max}$. Trustful data points are those of lower values compared to the corresponding $\alpha_{max}$, resulting in the cut-off frequencies indicated in (c). (e) and (f) the real and imaginary parts of the dielectric function, respectively. The error bars result from the standard deviation and thickness uncertainty.

## Note 3. Data from Experiment 2

Fig. S3 shows the data obtained from Experiment 2 at room temperature, following the same procedure of data analysis discussed in Note 2.

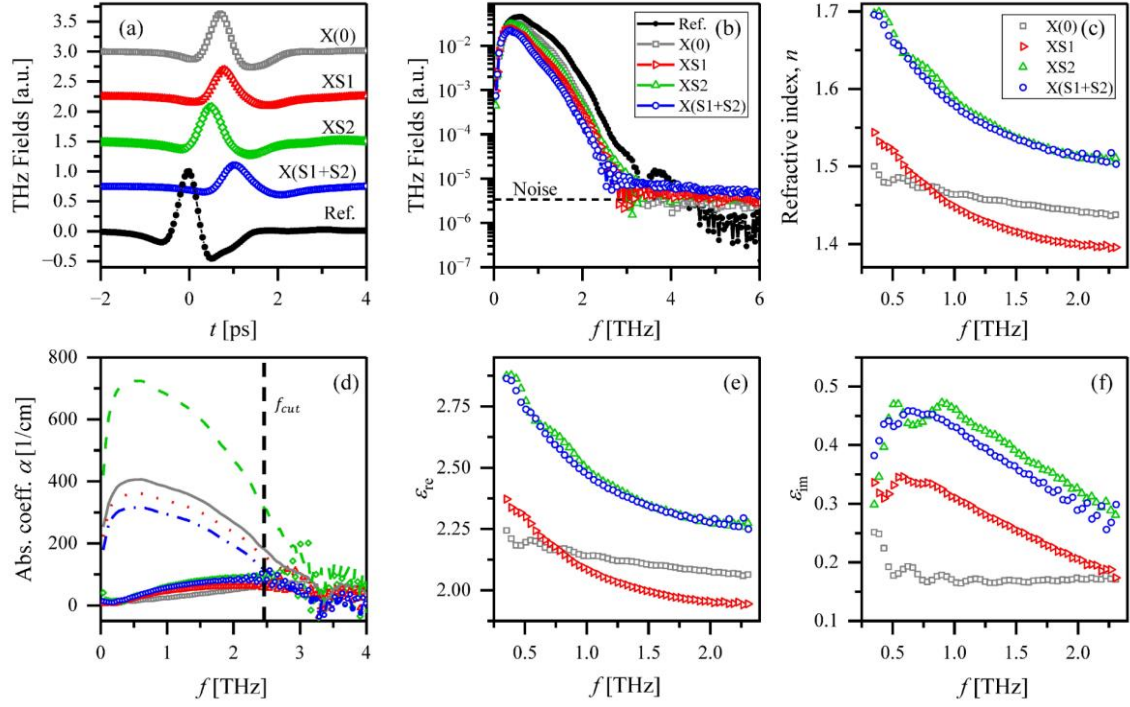

**Figure S3:** Data from Experiment 2. (a) the temporal waveforms of the THz signals transmitted through the samples and the reference signal transmitted through a sample-free aperture, (b) the corresponding amplitude spectra obtained by the fast Fourier transformation of the temporal signals of (a), (c) the real-valued refractive index. The arrows indicate the upper limit of the trustful spectral window for the trustful data for each sample. (d) the absorption coefficient, symbols for samples and dashed-lines for $\alpha_{max}$. Trustful data points are those of lower values compared to the corresponding $\alpha_{max}$, resulting in the cut-off frequencies indicated in (c). (e) and (f) the real and imaginary parts of the dielectric function, respectively.

## Note 4. Models of conductivity

To obtain deeper insights into the conduction properties of the electrolytes, the following conductivity models have been tested versus the data.

**Lorentz model of resonant vibrations**[10]:

$$\tilde{\sigma} = -i\varepsilon_0\omega(\varepsilon_\infty - 1) - \sum_{j=1}^{2} \frac{i\varepsilon_0\omega\Omega_j^2}{(\omega_{0,j}^2 - \omega^2) - i\gamma_j\omega} \quad (3)$$

where $\varepsilon_0$ is the vacuum permittivity, $\varepsilon_\infty$ is the high-frequency dielectric constant, and, for each mode of order $j$, $\Omega$ is the vibration amplitude (or oscillator strength), $\omega_0$ is the resonance angular frequency, $\gamma$ is the damping rate, and the sum is applied over the two modes (i.e. two terms with j=1,2) within the frequency range of our experiments.

**Table S1:** Fitting parameters used in the fitting with Lorentz model.

| Parameters | X(0) | XS1 | XS2 | X(S1+S2) |
|---|---|---|---|---|
| $f_{0,1}$ [THz] | 2.58 | 1.48 | 1.36 | 1.09 |
| $f_{0,2}$ [THz] | 4.56 | 5.43 | 5.12 | 6.04 |
| $\varepsilon_\infty$ | 1.93 | 1.88 | 2.11 | 2.06 |
| $\gamma_1$ [THz] | 35.49 | 25.32 | 20.17 | 16.19 |
| $\gamma_2$ [THz] | 22.93 | 41.60 | 44.31 | 62.97 |
| $\Omega_1/2\pi$ [THz] | 1.21 | 1.13 | 1.18 | 1.02 |
| $\Omega_2/2\pi$ [THz] | 1.51 | 1.91 | 2.88 | 4.08 |

**Debye model of dielectric relaxation**[11,12]:

$$\tilde{\sigma} = -(\varepsilon_\infty - 1) \cdot i\omega\varepsilon_0 - i\omega\varepsilon_0 \cdot \sum_{j=1}^{2} \frac{\varepsilon_j - \varepsilon_{j+1}}{1 - i\omega\tau_j} \quad (4)$$

The Lorentz conductivity model, Eq. (4), accounts for resonant modes of vibrations of bound polar molecules, where for each mode of order $j$, $\Omega$ is the vibration amplitude (or oscillator strength), $\omega_0$ the resonance angular frequency and $\gamma$ is the damping rate, and the sum is applied over the possible modes within the frequency range of our experiments. The fitting provided by Eq. (4) is represented by the solid lines in Fig. S4 (a and b), with the fitting parameters provided in Tab. S1.

Here $\varepsilon_1$ is the static dielectric constant (also known as $\varepsilon_s$), $\varepsilon_j$ are intermediate values of the dielectric function, and $\tau_D$ is the relaxation time. The fitting provided by Eq. (5) is represented by the solid lines in Fig. S4 (c and d), with the fitting parameters provided in Tab. S2. As reported previously[13], from the low frequency part of the spectrum (up to 2 THz) it is not possible to distinguish whether Debye or Lorentz model fits better, however, in the high frequency part of the broadband spectrum it is apparent that Lorentzian resonances reproduce the experimental data more precisely

**Table S2:** Fitting parameters used in the fitting with Debye model.

| Parameters | X(0) | XS1 | XS2 | X(S1+S2) |
|---|---|---|---|---|
| $\varepsilon_1$ | 2.11 | 2.64 | 3.23 | 3.44 |
| $\tau_1\ [fs]$ | 109.59 | 28.19 | 46.13 | 29.73 |
| $\varepsilon_2$ | 1.36 | 2.53 | 2.91 | 2.97 |
| $\tau_2\ [fs]$ | 201.37 | 301.37 | 277.72 | 339.58 |
| $\varepsilon_3\ (\varepsilon_\infty)$ | 1.87 | 1.81 | 2.01 | 1.87 |

**Combined Lorentz-Debye model:**

$$\tilde{\sigma} = -(\varepsilon_\infty - 1) \cdot i\omega\varepsilon_0 - i\omega\varepsilon_0 \cdot \frac{\varepsilon_s - \varepsilon_\infty}{1 - i\omega\tau_D} - \frac{i\omega\varepsilon_0\Omega^2}{(\omega_0 - \omega)^2 - i\omega\gamma} \quad (5)$$

This model is a combination of one Debye term (relaxation) and one Lorentzian resonance. The fitting provided by Eq. (5) is represented by the solid lines in Fig. S4 (e and f), with the fitting parameters provided in Tab. S3.

**Table S3**: Fitting parameters used in combined Lorentz-Debye model.

| Parameters | X(0) | XS1 | XS2 | X(S1+S2) |
|---|---|---|---|---|
| $\varepsilon_s$ | 2.22 | 2.58 | 3.05 | 3.17 |
| $\tau_D\ [fs]$ | 172.86 | 281.44 | 234.90 | 309.52 |
| $\varepsilon_\infty$ | 1.91 | 1.86 | 2.08 | 2.05 |
| $f_0\ [THz]$ | 4.25 | 4.95 | 4.75 | 5.52 |
| $\gamma\ [THz]$ | 21.11 | 27.70 | 29.45 | 42.22 |
| $\Omega/2\pi\ [THz]$ | 1.32 | 1.10 | 1.78 | 2.76 |

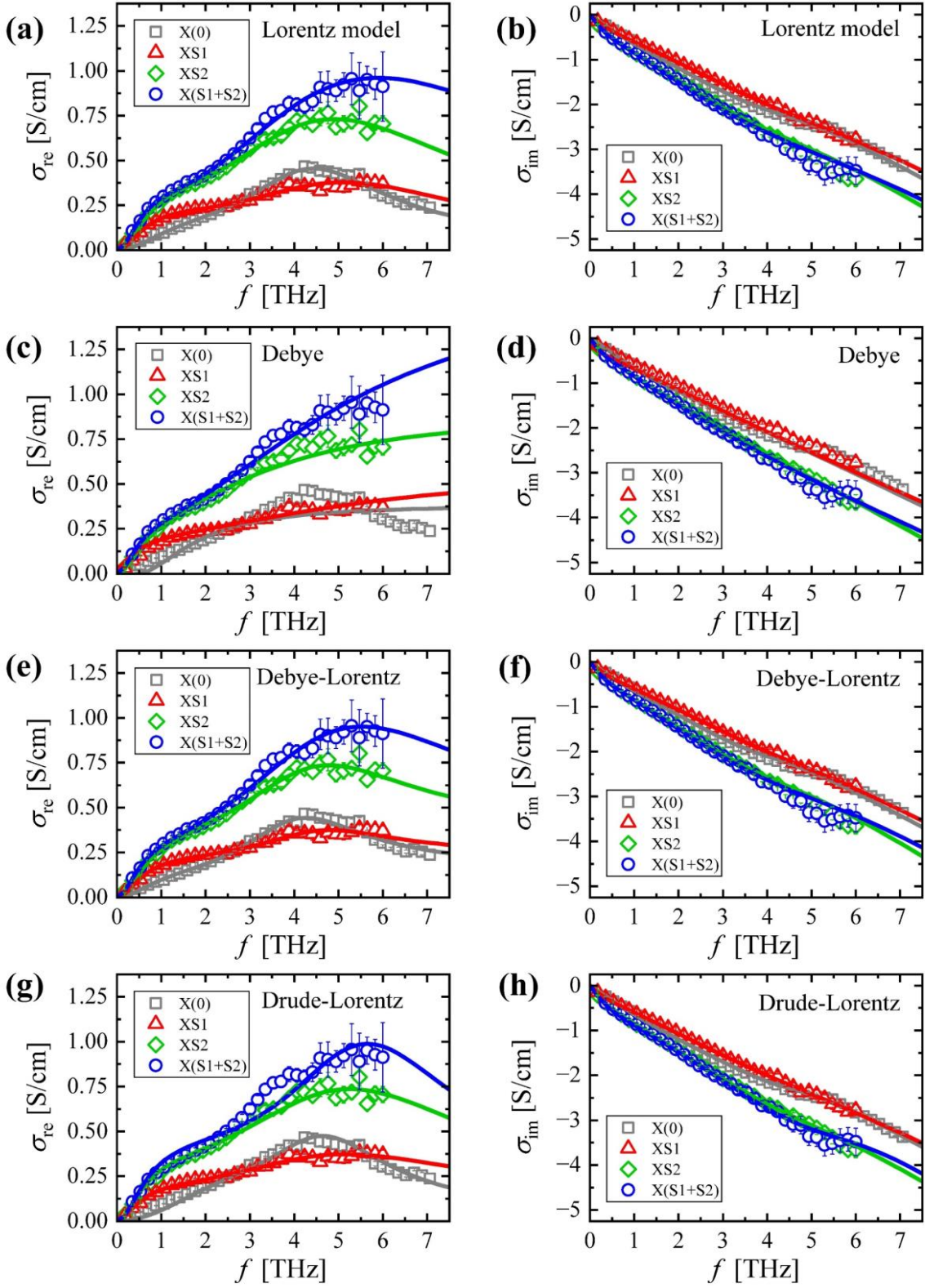

**Figure S4:** Fit results with various physical models: (a,b) Lorentz model using 2 Lorentzian resonances, (c,d) Debye model including 2 Debye terms, (e,f) a combined Debye-Lorentz model and (g,h) a model containing one Drude term and two Lorentzian resonances.

**Drude model of conduction by free charge carriers[6]:**

$$\tilde{\sigma} = \frac{\sigma_{dc}}{1 - i\omega\tau} \quad (6)$$

Here $\sigma_{dc}$ is the dc conductivity associated with the drift of free charge carriers driven by a dc electric field, and $\tau$ is the momentum scattering time. As shown in Fig. S4, the extrapolation of the data points of the real conductivity approaches zero at $f = 0$. This indicates that our samples exhibit a negligible $\sigma_{dc}$ as compared with the THz conductivity demonstrated in the frequency range of our experiments, and therefore the Drude model has no contribution in optimizing the fitting. This is also visible in the fit parameters of a combined Drude-Lorentz model which are displayed in Tab. S4 with the associated fit result shown in Fig. 2 (g and h), providing unrealistic extremely long ionic scattering times, $\tau_D$.

Table S4: Fitting parameters used in combined Lorentz-Drude model.

| Parameters | X(0) | XS1 | XS2 | X(S1+S2) |
|---|---|---|---|---|
| $\varepsilon_\infty$ | 1.92 | 1.88 | 2.13 | 2.12 |
| $\sigma_{dc}$ [S/m] | $5.46 \cdot 10^{-5}$ | $87.72 \cdot 10^{-5}$ | $5.90 \cdot 10^{-5}$ | $48.86 \cdot 10^{-5}$ |
| $\tau_D$ [ps] | 114.23 | 4822.36 | 2048.10 | 2146.58 |
| $f_{0,1}$ [THz] | 4.76 | 5.34 | 5.59 | 5.78 |
| $\gamma_1$ [THz] | 20.44 | 39.86 | 43.65 | 29.41 |
| $\Omega_1/2\pi$ [THz] | 1.41 | 1.48 | 2.65 | 2.38 |
| $f_{0,2}$ [THz] | 2.97 | 2.66 | 1.77 | 2.41 |
| $\gamma_2$ [THz] | 32.17 | 92.7 | 36.81 | 58.18 |
| $\Omega_2/2\pi$ [THz] | 1.29 | 2.27 | 1.76 | 2.56 |

**Note 5. Constant Fit Parameters from Temperature-Dependent Fitting**

For the analysis of the temperature-dependent THz-TDS measurements (experiment 2), $f_{0,2}, \Omega_1$ and $\Omega_2$ were held constant with temperature, due to the limited information given by the regarded spectral range. The associated parameters were determined by fitting the room-temperature data of experiment 2 together with the broadband data from experiment 1. The parameters are displayed in Tab. S5.

Table S5: Temperature-independent fitting parameters corresponding to the Lorentz model fit of experiment 2.

| Parameter | X(0) | XS1 | XS2 | X(S1+S2) |
|---|---|---|---|---|
| $f_{0,2}$ [THz] | 4.48 | 5.43 | 5.12 | 6.04 |
| $\Omega_1$ [THz] | 1.14 | 1.11 | 1.37 | 0.81 |
| $\Omega_2$ [THz] | 1.60 | 1.80 | 2.23 | 3.84 |

**Note 6. Aging and environmental effects**

To test the effects of humidity and aging on the properties of the electrolyte samples, we repeated the THz-TDS measurements and additionally performed Raman spectroscopy with the samples over one month of exposure to humid air. The results are displayed in Figs. S5 and S6, indicating almost no change in the properties of the samples.

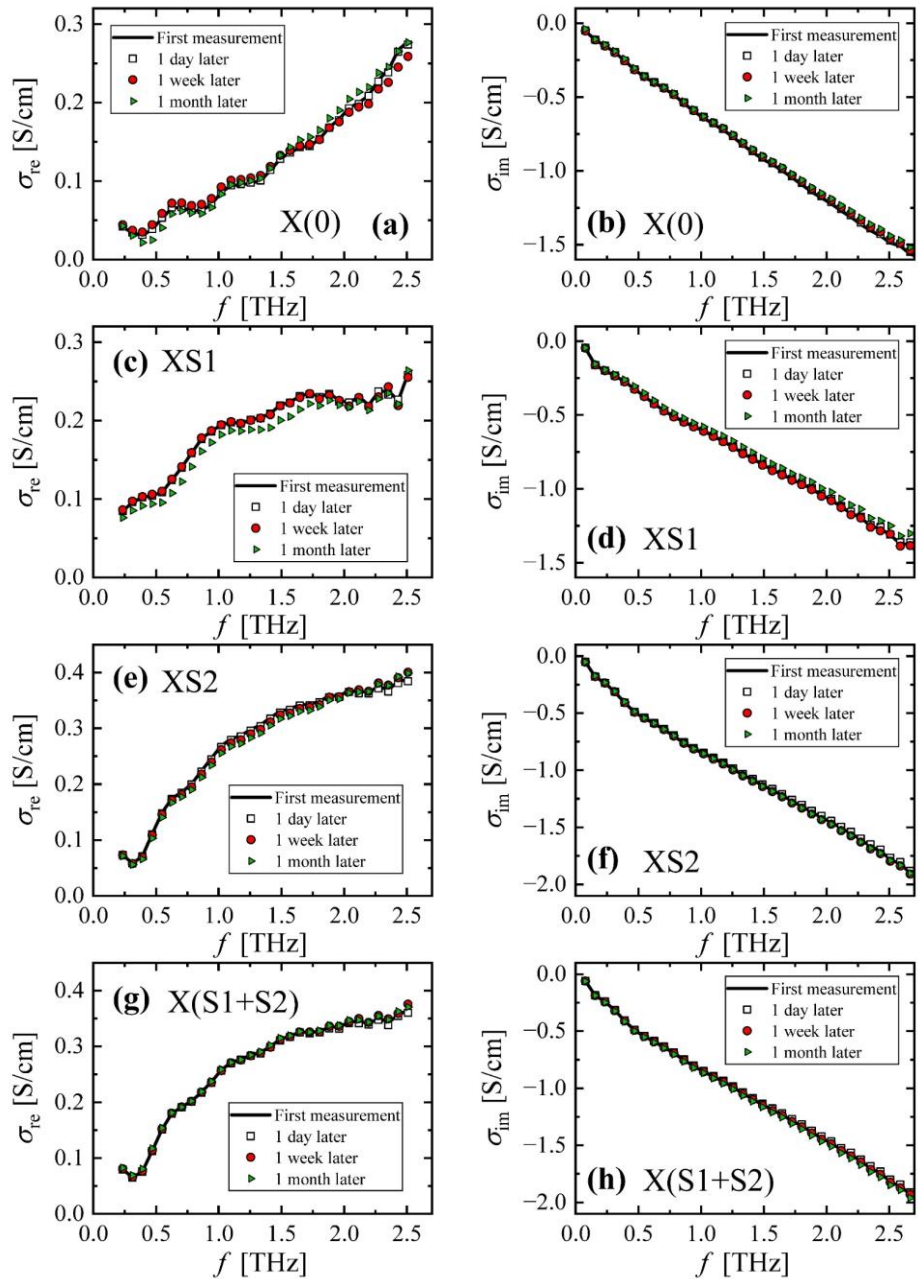

**Figure S5:** The conductivity spectra of polymer based electrolytes ((a,b) X(0), (c,d) XS1, (e,f) XS2, (g,h) X(S1+S2)) with the fresh samples (solid black line), after 1 day of exposure to humid air (open black squares), after 1 week of exposure to humid air (red circles) and after 1 month (green triangles).

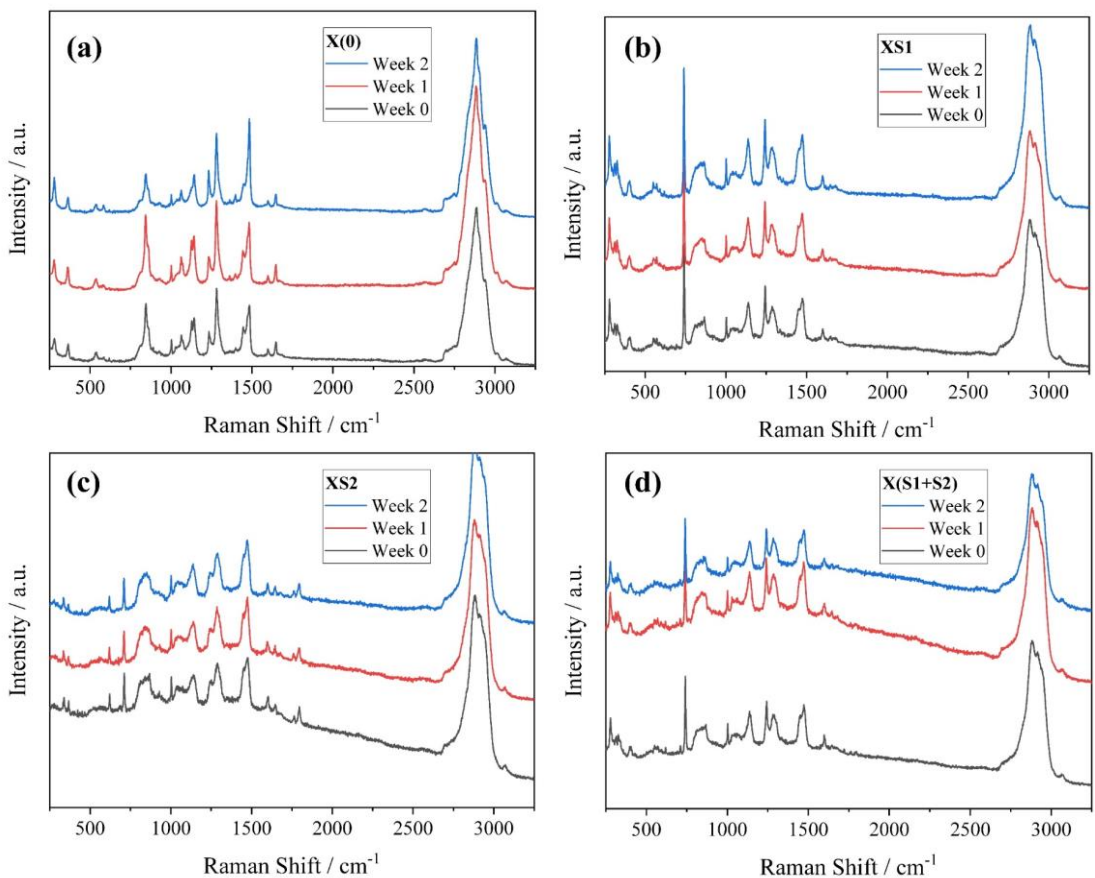

**Figure S6:** Raman spectra of the samples measured repeatedly at three different dates over three weeks, indicating no changes in the sample properties over the this period of time. (a) the bare polymer sample X(0), (b) and (c) the single-salt samples XS1 and XS2, respectively, and (d) the dual-salt sample X(S1+S2).

**Note 7. Measurements of the Ionic dc Conductivity of the SPEs**

Fig. S7(a) shows the electrically measured ionic conductivities of the SPEs XS1, XS2 and X(S1+S2) as functions of temperature ranging from 0 to 70°C. They were determined using Autolab instrument PGSTAT204-Fra32M, Metrohm Potentiostat) by electrochemical impedance spectroscopy (EIS) analysis. The impedance of the polymer membrane sandwiched between two stainless steel electrodes of a two electrode coin cell (CR-2032) was measured every 10 °C, after the given temperature was hold in equilibrium for 2 hours. From the impedance, the ionic conductivity was calculated using[14]

$$\sigma = lR_b/A \qquad (7)$$

Here, $l$ is the membrane thickness (in cm), $A$ the area of the sample and $R_b$ the resistance of the bulk (the equivalent circuit model is shown in Fig. S8). In the measurement, the frequency range of the AC perturbation was chosen between 200 KHz and 1Hz, with an amplitude of 10 mV. In Fig. S7(a), it is visible that while XS1 shows a comparatively low ionic conductivity, XS2 and X(S1+S2) show very similar values which approach each other for lower temperatures. In Fig. S7(b), it is shown that the ionic conductivity of the dual salt electrolyte is enhanced further by increasing the LiDFOB concentration slightly but decreases for even higher LiDFOB concentrations. A similar phenomenon is observable in Fig. S7(c) where the temperature-dependent ionic conductivity of single salt electrolytes containing different concentrations of LiTFSI is shown. Here, the maximum conductivity is obtained by adding 20 wt% of the salt. For clarification, in Table S6, the room temperature ionic conductivities of various samples containing different salt compositions are listed.

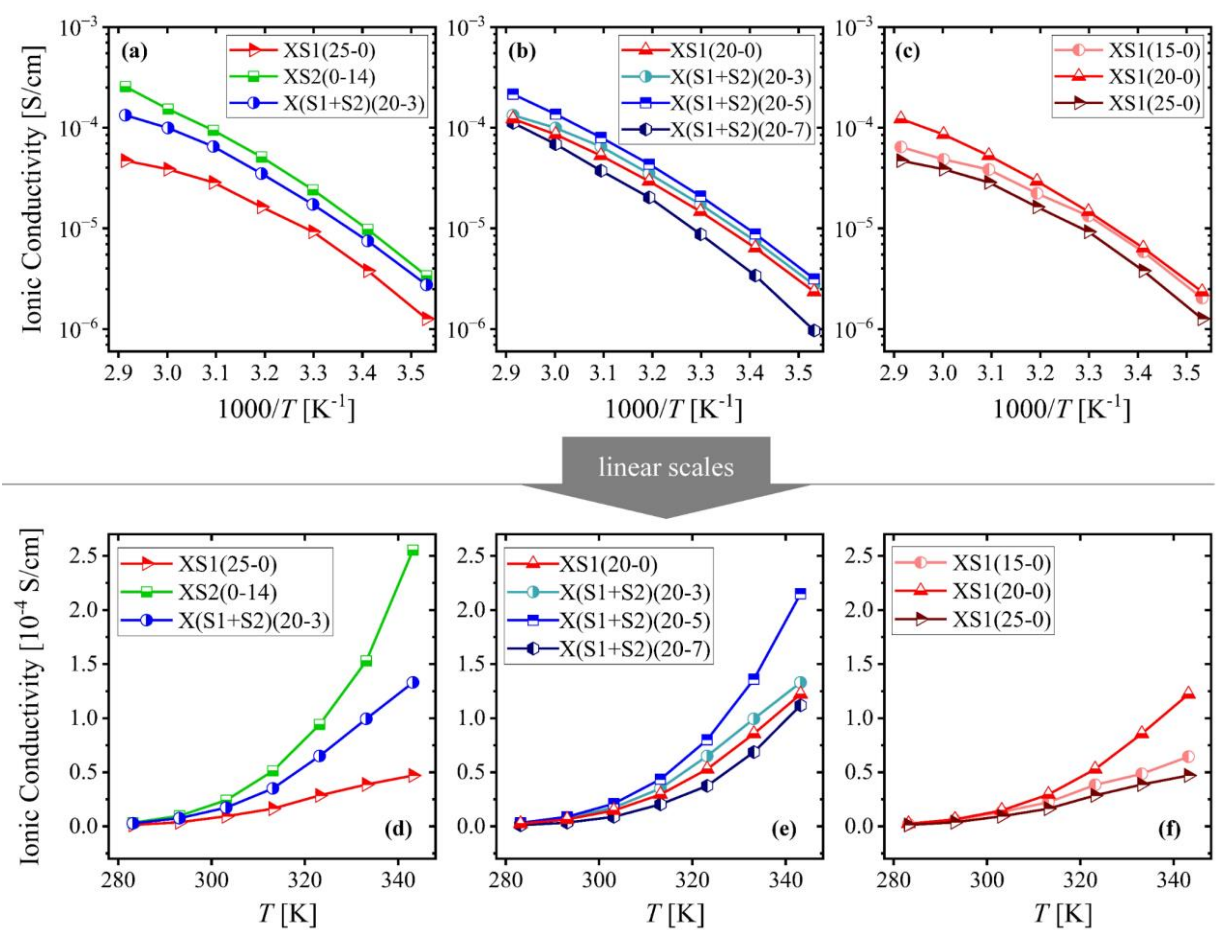

**Figure S7:** (a) Results of temperature-dependent electronical ionic conductivity measurements of the SPEs XS1, XS2 and X(S1+S2) containing the same concentrations of salts as used in the THz measurements (XS1 – 25 wt% LiTFSI, XS2 – 14 wt% LiDFOB, X(S1+S2) – a mixture of 20 wt% LiTFSI and 3 wt% LiDFOB). (b) Ionic conductivities of dual salt electrolytes containing different concentrations of LiDFOB, compared to the sample XS1. Apparently, the maximum ionic conductivity values are obtained by the sample containing 20 wt% LiTFSI and 5 wt% LiDFOB. (c) Ionic conductivities of single salt electrolytes containing different concentrations of LiTFSI. (d-f) The same data as presented in (a-c) but in linear scaling for comparability to the main text.

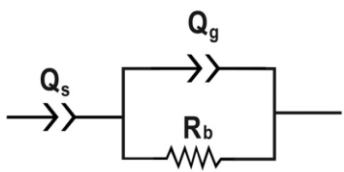

**Figure S8:** Circuit model used to fit the EIS measurements for ionic conductivity.

**Table S6**: Ionic conductivities of X(0), XS1, XS2, X(S1+S2) measured by EIS at 293 K.

|  | XS1 | XS1 | XS1 | XS2 | X(S1+S2) | X(S1+S2) | X(S1+S2) |
|---|---|---|---|---|---|---|---|
| **Salt concentration (LiTFSi-LiDFOB) [wt%]** | (15-0) | (20-0) | (25-0) | (0-14) | (20-3) | (20-5) | (20-7) |
| **Ionic conductivity [S/cm] at 293 K** | $5.85 \cdot 10^{-6}$ | $6.39 \cdot 10^{-6}$ | $3.80 \cdot 10^{-6}$ | $9.76 \cdot 10^{-6}$ | $7.5 \cdot 10^{-6}$ | $8.83 \cdot 10^{-6}$ | $3.41 \cdot 10^{-6}$ |

We note here that the bare PEO is neither electrolyte nor electrically conductive (i.e. it is an insulator exhibiting no measurable ionic conductivity). When a salt is added to the PEO to form an electrolyte, ionic charge carriers are created in the formed electrolyte and an ionic conductivity evolves and increases with the addition of salt until a maximum conductivity is reached. Further addition of salt will then lead to a reduction of the electrolyte ionic conductivity, demonstrated in consistency with the study by Wu et al.[15], which has been attributed to an increase of the insolubility of the comprised salts in the polymer matrix [14,16]. By blending different types of salt, the salt concentration yielding the maximum ionic conductivity can be shifted to higher levels[17].